\documentclass[twocolumn,secnumarabic,amssymb, nobibnotes, aps, prd]{revtex4-2}

\usepackage{graphicx, bm}
\begin{document}

\title{
Period and intrinsic noises of permanent spin vortex system centered by small polaron
}%

\author{Hikaru Wakaura}%
\email[Quantscape: ]{
hikaruwakaura@gmail.com}
\affiliation{QuantScape Inc. QuantScape Inc., 4-11-18, Manshon-Shimizudai, Meguro, Tokyo, 153-0064, Japan}

\author{Takao Tomono}

\affiliation{ Digital Innovation Div. TOPPAN Inc. , 1-5-1, Taito, Taito, Tokyo, 110-8560, Japan}
\email[TOPPAN: ]{takao.tomono@ieee.org}
\date{December 2018}%

\begin{abstract}
Some fascinating phenomena have been investigated, hence spin vortex systems are a hot subarea of solid-state physics. However, simulations of these systems require some simplifications. Quantum computers have the potential to simulate these systems without them. Therefore, we simulated the time propagation of spin vortex systems centered by small polarons by the simulator of quantum computers. As a result, we revealed that these systems have much shorter periods than XXZ models and are never be relaxed. Though there is intrinsic persistent noise, this result will affect both fundamental and applied solid-state physics.

\end{abstract}

\maketitle
\tableofcontents
\section{Introduction}\label{1}
The movement of simulating quantum systems occurred in the 80s' by the development of classical computers\cite{feynman_simulating_1982}. It started from the simulation of the Ising model and gases of fermion and bosons. Today, reactions of proteins can be simulated by supercomputers\cite{YANG2016169}.
Quantum simulators that can simulate the quantum systems have also been developed such as ultracold atoms\cite{Bloch2012QuantumSW}, Ising machines\cite{d-wave2011}\cite{Utsunomiya:11} and Rydberg atoms\cite{2020JPhB...53a2002A}. Cold atoms and Rydberg atoms can simulate the Bose-Einstein condensation\cite{perrin_ultra_2009}\cite{2012PhRvL.109k5301Z} and many-body interactions\cite{2015arXiv151006403Y}\cite{Samajdare2015785118}. Ising machines can be used for simulations of phase transitions\cite{2021arXiv211005124P} and quantum annealing\cite{2020arXiv201109495G}. However, quantum simulators are not available for users outside of their lab except the Ising machines.

In parallel to the development, quantum hardwares have been improved concerning both the number of qubits and quantum volume. Ion-trap quantum computers have been developed rapidly for last five years. They suppass the quantum computers of superconductivity qubits with respect of both the number of qubits and quantum volume.  Both Honeywell \cite{2020PhRvR...2a3317B} and Ion-Q \cite{IonQ2020} updated the world record of quantum volume twice and the record of this is four million achieved by the quantum computer of Ion-Q. Other qubits have also been developed such as photonic qubits\cite{Zhong1460} and silicon qubits.

Today, quantum hardwares can be used for quantum simulation. Recently, the movement of simulation of quantum computers has been accelerated around the world. For example, time propagation on quantum open system has been simulated\cite{PhysRevD.104.L051501}\cite{2021PhRvL.127b0504H}. Quantum phase transition has been simulated by the quantum Monte-Carlo method. Quantum neural network has revealed that can be used for quantum simulation\cite{2019PhRvL.122y0501N}. Large-scale quantum systems cannot be simulated yet, quantum simulation on quantum computers has the potentials to simulate the systems without simplifications we use in those on classical computers.

Spin vortices centered by small polarons are one of the systems that Noisy-Intermediate-Scale-Quantum(NISQ) computers and their simulators can simulate. Pair spin vortices of Melon and Anti-melon vortices emerge in the 2D plane of fermions lattices. This is the Kostaritz-Thouress transition. Spin-vortex systems are have been researched investigated by many groups around the world. For example, topological excitations induced by spin vortex\cite{wang_evidence_2021} and time propagation of spin waves\cite{PhysRevLett.109.133602} are have been studied by quantum simulators. Spin vortex systems are also promising concerning an application for devices. Graphene nanosheets can be used as catalysts\cite{kwon_extremely_2018} by doping holes and persistent currents induced by spin vortices of itinerant electrons called spin-vortex-induced loop currents (SVILCs) are predicted theroretically\cite{koizumi_persistent_2014}. In addition, SVILCs are predicted that can be used for quantum computers\cite{Wakaura201655}.
Therefore, we simulated the time propagation of spin vortices centered by small polarons and calculated the states and magnetic moments by the Suzuki-Trotter manner. As a result, we find that there is periodic intrinsic magnetic noise.

The following organizations of this paper are as follows. Chapter \ref{2} is describing the detail of our method. Chapter \ref{3} indicates the result of our simulation. Chapter \ref{4} is the conclusion of our works.

\section{Method}\label{2}
\begin{figure}[h]
\includegraphics[scale=0.25]{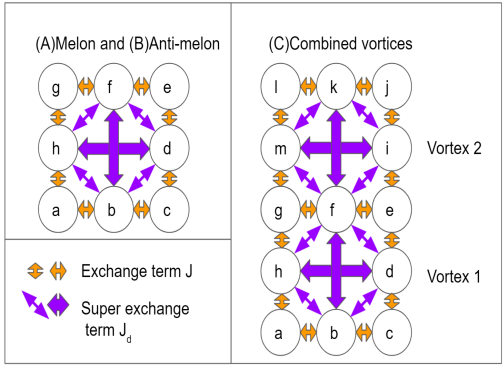}
\caption{
Numbers of sites of (A)Melon vortex, (B)Anti-melon vortex and (C)Combimed vortices, respectivly.
}\label{refsv}
\end{figure}

\begin{figure}[h]
\includegraphics[scale=0.25]{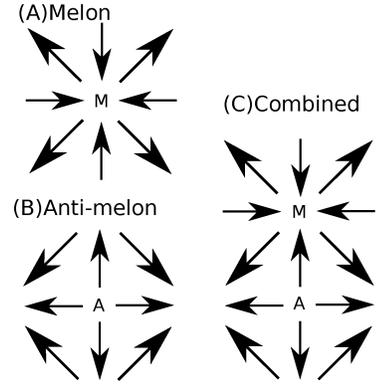}
\caption{
Spin structure of (A)Melon vortex, (B)Anti-melon vortex and (C)Combimed vortices, respectivly.
}\label{svst}
\end{figure}

\begin{figure}[h]
\includegraphics[scale=0.7]{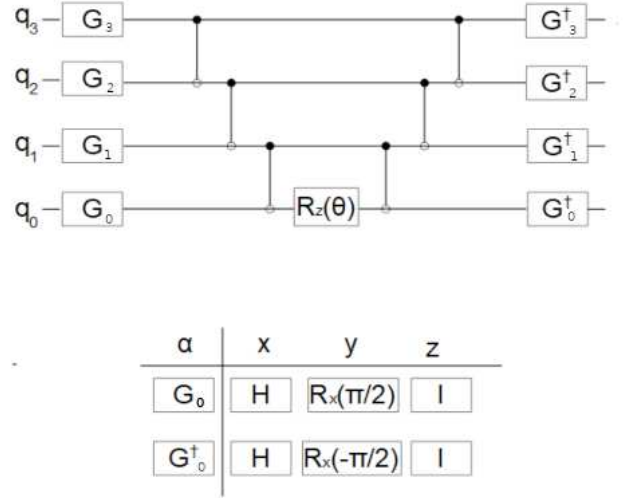}
\caption{The quantum circuit that is used for the VQE method that performs $exp(-i\theta\sigma_0^\alpha\sigma_1^\beta\sigma_2^\gamma\sigma_3^\delta)$. $G_j$ gate is $H$, $R_x(\pi/2)$ or $I$ gate depend on what pauli gate is operated for $q_j$.}\label{vqeexp}
\end{figure}

In this section, we describe our method for quantum simulation and its detail. We describe the Hamiltonian of spin vortex system as spin hamiltonian including exchange, and superexchange terms by the following equation.

\begin{equation}
H=\sum_{\langle p, q \rangle}^{N}\bm{S}_p\bm{S}_q+\sum_{\langle p, q \rangle_h}{S}_p\bm{S}_q,
\end{equation}

where $\bm{S}_p$ indicates the spin vector that satisfies $\bm{S}_p=(cos\xi_p sin\theta\_p \sigma_p^x, sin\xi_p sin\theta_p \sigma_p^y, cos\theta_p \sigma_p^z)$ and $\langle p, q \rangle$ indicates the nearest neighbor sites, and $\langle p, q \rangle_h$ indicates the next nearest neighbor and sites that across the hole like Fig. \ref{refsv}. Spin vortices have winding numbers that distinguish the direction of vortex, Melon, and Anti-melon. $J$ is $2t^2/U$ for $t=0.13eV$ and $U=8t$ that is 4$^{th}$ perturbation of Hubbard hamiltonian's energy. These terms are those of cuprate superconductivities\cite{HKoizumi2013}. We demonstrate the time propagation of (A)Melon vortex, (B)Anti-melon vortex, and (C)Combined vortices as shown in Fig. \ref{svst}. $\theta$ is $\pi/2$ for all sites, thus, all spins lie on the XY plane. We had better calculate including kinetic and Coulomb terms for more realistic simulation. Although, this makes; Hamiltonian more complicated. Therefore, we simulate only on spin Hamiltonian. We simulate the time propagation by acting $e^{-iH\Delta t}$ on the initial state. However, this Hamiltonian cannot be processed by quantum computer without some techniques\cite{McClean_2016}. Time propagator $e^{-iH \Delta t}$ must be decomposed into the product of single terms of Hamiltonian $H=\sum_j H_j$ as $e^{-iH\Delta t}=lim_{N\rightarrow \infty}(\prod_je^{-iH_j\Delta t/N})^N$ by Suzuki-Trotter decomposition method\cite{2019arXiv190401336J}. All terms can be expressed as a product of time-promoting operations, and a quantum computer can handle them as gate operations. The circuit would look like Fig. \ref{vqeexp}.


This circuit operates on terms consisting of multiple Pauli operators in a four-qubit system. The $q_0,q_1,q_2$, and $q_3$ represent the first, second, third and fourth qubits, respectively. $q_0,q_1$ represent the coupled orbits, and $q_2,q_3$ represent the anti-coupled orbits, respectively. The $R_z(\theta)$ exists to multiply the coefficients, and the $G_j$ and $G_j^\dagger$ are determined by which the Pauli operator is multiplied to the $j$-th bit.
For example, when multiplying $exp(-i\theta_k\sigma_0^x\sigma_1^y\sigma_2^z)$, $G_0$ is the $H$ gate, $G_1$ is $R_x(\pi/2)$, and $G_2$ is the $I$ gate (unit gate). Since there is no Pauli operator acting on $q_3$, there is no CNOT between $G_3$ and $q_2,q_3$. Also, when multiplying by $exp(-i\theta_k\sigma_0^x\sigma_2^y\sigma_3^z)$, the CNOT gate between $q_0$ and $q_1$ replaces the gate between $q_0$ and $q_2$, the CNOT gate between $q_1$ and $q_2$ and $G_2$ disappear, and $G_3$ and $G_3^\dagger$ are no longer in use. The CNOT gate between $q_1$ and $q_2$ and $G_2$ disappear, and $G_3$ takes the place of $G_2$. Everything else is processed in same way. In this paprer, the depth of time propagator (number of repetation of propagation on single terms) is 1. We assume that $\Delta t$ is constant and perfome time propagation for $nT$ seconds for period $T$. $T$ is $40.5054(fs)$ calculated by exchange term and magnetic dipole moment of electron. All datas are sampled by blueqat SDK\cite{blueqat}, which is the simulater of quantum computers. All results are the form of statevector(the number of shots is infinity).
Then, $\bm{S}_p=(sin\theta_p cos\xi_p \sigma_p^x, sin \theta_p sin \xi_p \sigma_p^y, cos \theta_p \sigma_p^z)$.

\section{Result}\label{3}
\begin{figure*}[h]
\includegraphics[scale=0.5]{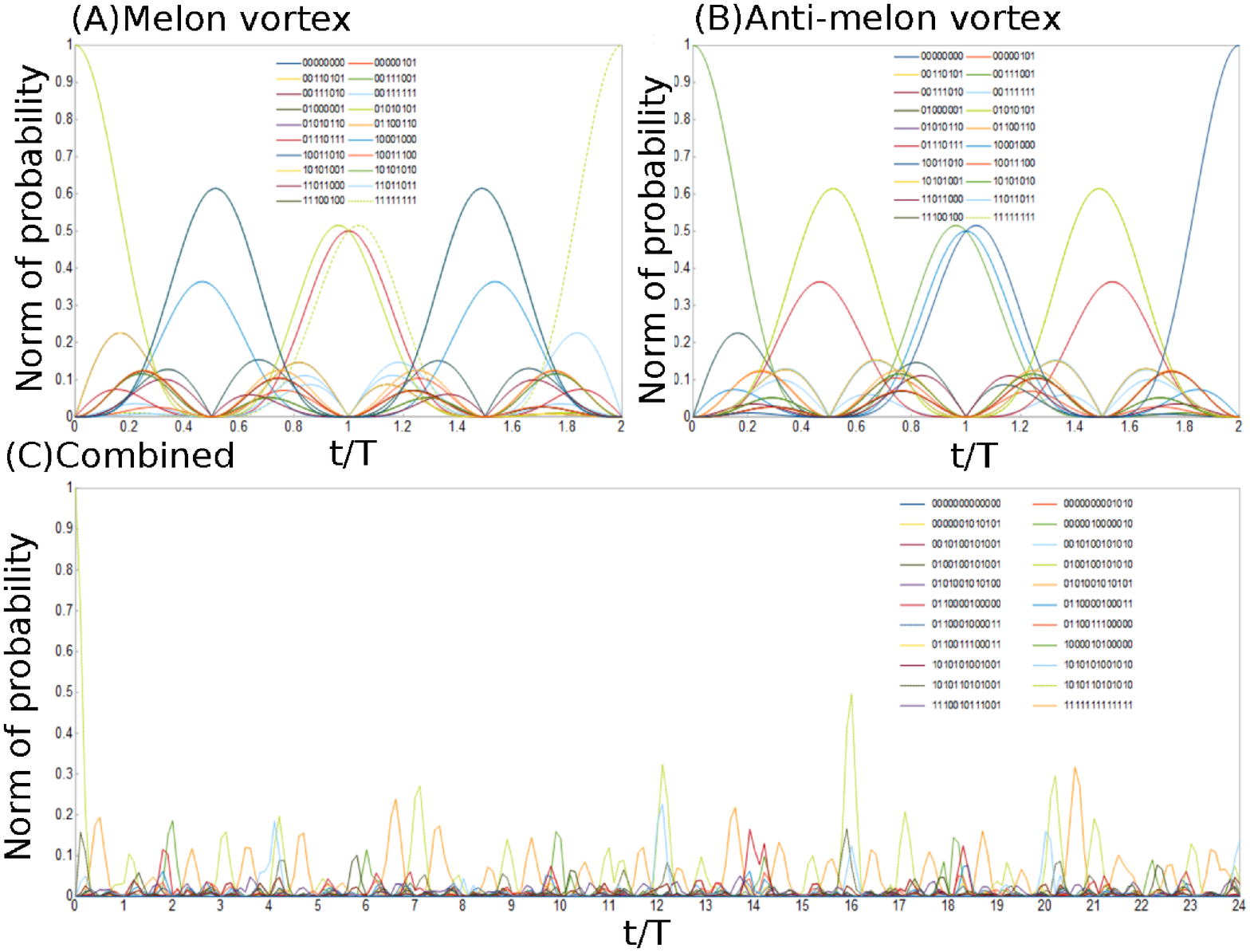}
\caption{
Time propagations on states that the norm of coefficients varies prominently of (A)Melon vortex, (B)Anti-melon vortex and (C)Combimed vortices, respectivly. $\Delta t$ is T/300, T/300 and T/10(s) for (A), (B) and (C), respecvtively.
}\label{svtp}
\end{figure*}

\begin{figure*}[h]
\includegraphics[scale=0.5]{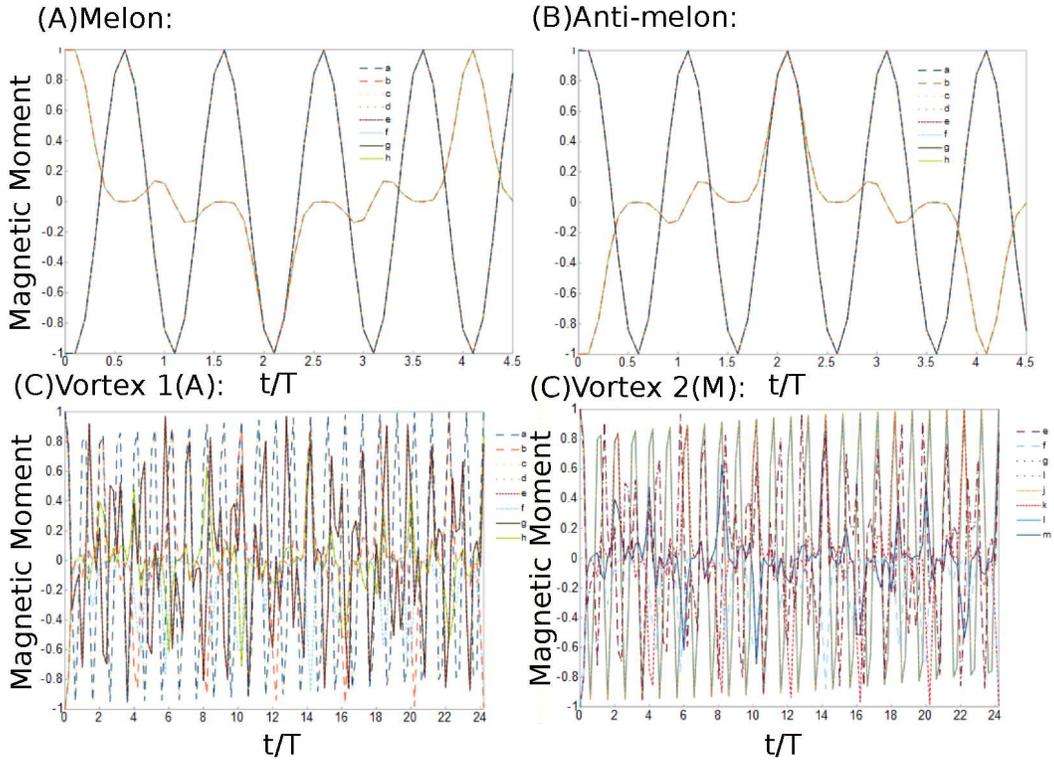}
\caption{
Time propagations on z component of magnetic moments on each site of (A)Melon vortex, (B)Anti-melon vortex, and (C)Combined vortices, respectively. $\Delta t$ is T/300, T/300, and T/10(s) for (A), (B), and (C), respectively. They are sampled from zero to end in $20\Delta t$ pitch.
}\label{svtm}
\end{figure*}

In this section, we describe the time-propagation of states and energies of spin-vortex systems. States are described as the z component of spins on each site expressed as $\mid mlkjihgfedcba \rangle$. Firstly, we show the time-propagation of states of (A)Melon vortex, (B)Anti-melon vortex, and (C)Combined vortices. Initial state of systems are $\mid 10101010 \rangle$, $\mid 01010101 \rangle$ and $\mid 0101010110101 \rangle$, respectively. Note that $0$ and $1$ are corresponding to up spin and down spin, respectively. According to Fig. \ref{svtp}, initial states of (A)Melon vortex and (B)Anti-melon vortex decrease and flipped state and the states that all spins are same increase the norm of coefficient when $t/T\leq 2.0$. All square roots of probability are symmetrical for $t/T=2.0$, hence time propagation of all states has the period of $4T$. In contrast, the initial state of the combined spin-vortex decreases rapidly and the state that all spins are the same never increases the probability to 1 in sampled timespan. In exchange, the coefficients of some states increase frequently.

the z component of Magnetic moment of each site of (A)Melon vortex, (B)Anti-melon vortex, and (C)Combined spin vortex are shown in Fig. \ref{svtm}. Magnetic moment of site x is calculated by $\langle \Phi \mid \sigma_x^z \mid \Phi \rangle=M_x^z$, hence they are from -1 to 1. Magnetic moments on vertex sites of Melon and Anti-melon vortex propagate identically. Those online sites are also. Magnetic moments on all sites spike at $t=2T$. At this moment, a temporary magnetic pulse occurs on the systems. Besides, the direction of it depends on the initial state. However, the sign never depends on the winding number.
In addition, x and y components are zero for all sites because they are nullified by the superposition of states. Magnetic moments of (C)Combined vortices propagate in symmetry for the center cite f. The magnetic moment on each vortex propagates identically.
For example, site a, c, j, and l propagates are the same for all timespan. The time propagation of site b and k are also the same. The time propagation of d, h, i, and m are also. At 24T, The magnetic moments on all sites become near the initial state and the maximum of the site a, c, j, and l near to 1 for this timespan. Hence, the period of (C)Combined vortices can be estimated to be the multiplied number of 24T. The result of semiclassical simulations that uses statevectors of previous states and act propagator on them shows that the period of (C)Combined vortices is 48T as shown in Table. \ref{tabsv}.

\begin{table}[h]
\caption{
Estimated period of XXZ systems, (A),(B)Single vortex and (C)Combined vortices estimated by semiclassical calculation that uses statevectors as initial values and act propagator only once.
}\label{tabsv}
\begin{tabular}{c|c}\hline \hline
system&period(T) \\ \hline
XXZ,$\Delta=0$&$\geq$ 400 \\
XXZ,$\Delta=2$&$\geq$ 400 \\
(A),(B)Single vortex& 4 \\
(C)Combined vortices&48 \\ \hline
\end{tabular}
\end{table}

Ordinary spin systems relax a long time after and integrable spin systems such as XXZ systems\cite{PhysRevX.11.011011} have much longer periods than spin vortex systems. As shown in Table. \ref{tabsv}, the period of XXZ system is more than 400T regardless of polarization of z spins. It is supposed to be due to the many-body localization\cite{2021arXiv210913608S} occurred by small polarons and superexchange terms around them. However, the accuracy of simulation may be insfficient for the simulation of  for long periods. Chebychev manner that is well-known methods for simulation of time propagation will improve the accuracy of the simulations on quantum computers\cite{2021arXiv210913608S}.

\begin{figure*}[h]
\includegraphics[scale=0.5]{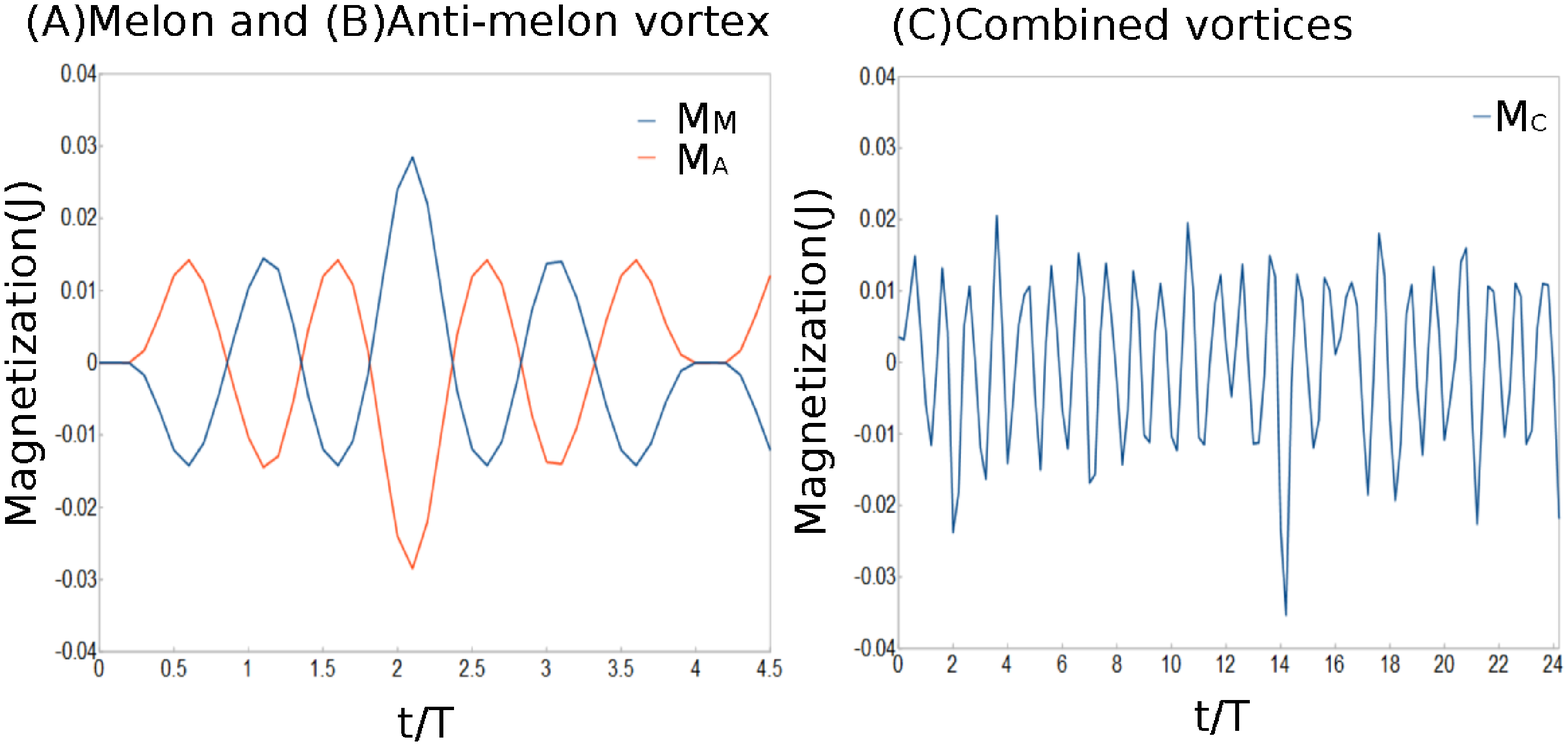}
\caption{
Time propagations on magnetization in a unit of exchange integral J on each site of (A)Melon vortex, (B)Anti-melon vortex, and (C)Combined vortices, respectively. $\Delta t$ is T/300, T/300, and T/10(s) for (A), (B), and (C), respectively. They are sampled from zero to end in $20\Delta t$ pitch for (A), (B) and $2\Delta t$ for (C), respectively. $M_M$, $M_A$, and $M_C$ are magnetizations of (A)Melon vortex, (B)Anti-melon vortex, and (C)Combined vortices, respectively.
}\label{svinm}
\end{figure*}

Spin vortex systems have intrinsic magnetic noises compared to other spin systems.  They are induced by spin vortices themselves. Although, temporal pulses harm SVILCs because the time average of this current is zero for the period propagation. These pulses can be named spin-vortex-induced noise moments (SVINMs). SVINM fluctuates the energy levels in sub-picoseconds order time through Zeeman splitting and magnetic dipole moment. Besides, this moment has a magnitude of 3.5662$\times$10$^{-3}$J/T for exchange integral J. We show the SVINMs of (A)Melon vortex, (B)Anti-Melon vortex, (C)Combined vortices in Fig. \ref{svinm}. As the number of vortices increases, the time propagation of SVINMs becomes more complicated.

The energy levels of each system are 0 for all times because diagonal terms of Hamiltonian are all zero. The thermal average of energy in the microcanonical ensemble is equal to energy when $\beta=0$ in all cases. Although, any other eigenstates never match it by only time propagation. Hence, these spin vortex systems satisfy only weak ETH.

\section{Concluding remarks}\label{4}
Spin vortex propagates and the time propagation of them has the period. This means Zeeman splitting is nonnegligible for all spin vortex systems. Such systems inevitably suffer from intrinsic magnetic noises called SVINMs as the intensity of the environmental magnetic field increases. Many systems take advantage of spin vortices such as SVILCs and nano-sheet of graphene. Especially for SVILCs, Zeeman splitting of current modes is necessary for quantum computation. Alternative manners are required to circumvent the SVINMs such as external currents for splitting not only for coupling\cite{Wakaura201655}. Rashba spin-orbit coupling\cite{Koizumi2017} may relieve the SVINMs. Next problem is to take into account it for our simulations. However, spin vortex systems have been revealed that never be relaxed without external noises and have short periods because of small polarons. This will foster the applications of these systems for devices.

\newpage\bibliographystyle{apsrev4-2}
\bibliography{temp7}

\end{document}